# Observational evidence for Extended Emission to GW170817

Maurice H.P.M. van Putten[1] and Massimo Della Valle[2]
[1] *Physics and Astronomy, Sejong University, 98 Gunja-Dong Gwangin-gu, Seoul 143-747, Korea*
[2] *Ist. Nazionale di Astrofisica, Osservatorio Astronomico di Capodimonte, Napoli, Italy, Inst. Astrofisica de Andalucia, Granada, Spain*

**ABSTRACT**
The recent LIGO event GW170817 is the merger of a double neutron star system with an associated short GRB170817A with $2.9 \pm 0.3$ s soft emission over 8-70 keV. This association has a Gaussian equivalent level of confidence of $5.1\sigma$. The merger produced a hyper-massive neutron star of stellar mass black hole with prompt or continuous energy output powering GRB170817A. Here, we report on a possible detection of Extended Emission (EE) in gravitational radiation *during* GRB170817A: a descending chirp with characteristic time scale $\tau_s = 3.01 \pm 0.2$ s in a (H1,L1)-spectrogram up to 700 Hz with Gaussian equivalent level of confidence greater than $3.3\sigma$ based on causality alone following edge detection applied to (H1,L1)-spectrograms merged by frequency coincidences. Additional confidence derives from the strength of this EE. The observed frequencies below 1kHz indicate a hyper-massive magnetar rather than a black hole, spinning down by magnetic winds and interactions with dynamical mass ejecta.

**Key words:**  galaxy dynamics: observations

## 1 INTRODUCTION

The LIGO-Virgo detection of GW170817 (Abbott et al. 2017a) followed by the *Fermi*-GBM and INTEGRAL identification GRB170817A (Connaughton 2017; Savchencko et al. 2017) for the first time establishes a merger as the progenitor of a GRB. GRB170817A is classified as a short gamma-ray burst (SGRB) at $T_{90} = 2.0 \pm 0.5$ s ($\leqslant 300$keV) (Goldstein et al. 2017) with relatively soft extended emission over $T_{90} = 2.9 \pm 0.3$ ($\leqslant 70$keV) (Pozanenko et al. 2018) whose isotropic equivalent energy is below that of typical SGRBs by about four orders of magnitude (Kasen et al. 1017).

The GW170817/GRB170817A association is significant for our understanding of the origin of heavy elements by the r-process (Kasen et al. 1017; Smartt et al. 2017; Pian et al. 2017) and it possibly provides new measurements of the Hubble parameter $H_0$, independent of any of the existing methods Abbott et al. (2017); Guidorzi et al. (2017); Riess et al. (2018)/ Observing tens of neutron star mergers in the near future may resolve the $H_0$ tension problem (Freedman 2017), promising new insights on weak gravity at the de Sitter scale of acceleration in the Universe (van Putten 2017b).

While GW170817 establishes a double neutron star merger as the origin of GRB170817A, it leaves unidentified its central engine (Usov 1992): is GRB170817A powered by prompt *or* continuous energy output from a newly formed hyper-massive neutron star *or* black hole (Usov 1992; Woosley 1993; Nakar 2007; Ruffini et al. 2016a,b; Nagataki 2018)? It is widely believed that gravitational-wave observations have the power to resolve these questions (Cutler & Thorne 2002).

GW170817 produced a clear signal over 40-300 Hz in the Hanford (H1) and Livingston (L1) detectors of LIGO by an energy output $E_{GW} > 2.5\% \, M_\odot c^2$, where $c$ is the velocity of light. This may be accompanied by gravitational-wave emission just prior and after final coalescence. For instance, high frequency emission is expected from tidal effects starting at $f_t \sim 600$ Hz in the run-up to final coalescence (Damour et al. 2012; Del Pozzo et al. 2013; Hinderer et al. 2009; Reed et al. 2009; Wade et al. 2014), possibly accompanied by dynamical mass ejections (Baiotti et al. 2008; Faber & Rasio 2012; Baiotti & Rezzolla 2017) and post-merger signals may derive from a rapidly spinning central engine powering GRB170817A. Some confidence in this outlook derives from high-resolution numerical simulations, showing the formation of hyper-massive neutron star-disk systems soon after $f_t$. After some delay, this system may collapse to stellar mass black hole (Baiotti et al. 2008), as in core-collapse of massive stars, e.g., SN1987A (Burrows & Lattimer 1987; Brown et al. 2003).

However, rigorous identification of the central engine of GRB170817A defies electromagnetic observations, given significant uncertainties in circumburst environment, viewing angle and the unusual spectral-energy properties of this event (Abbott et al. 2017b; Pozanenko et al. 2018). GW170817A may have formed a long-lived hyper-massive neutron star, similar to the high temperature proto-neutron star in SN1987A inferred from neutrino energies in excess of 10MeV. Its ~ 10 s light curve is commonly attributed to slow neutrino diffusion in matter at supra-nuclear densities (Brown et al. 2003). but might also derive from a high-density accretion disk around a newly formed black hole derived from fall back of a remnant stellar envelope giving $L_{\bar{\nu}_e} \simeq 10^{51}\,\mathrm{erg\,s^{-1}}$ after the first few seconds (Blum & Kushnir 2016). Final remnants appear to be black holes, inferred from late-time X-ray emission (Ruan et al. 2018). If detected, an Extended Emission in gravitational-wave emission potentially provides a direct signature of a central en-gine, circumventing aforementioned uncertainties in electromagnetic and neutrino emission.

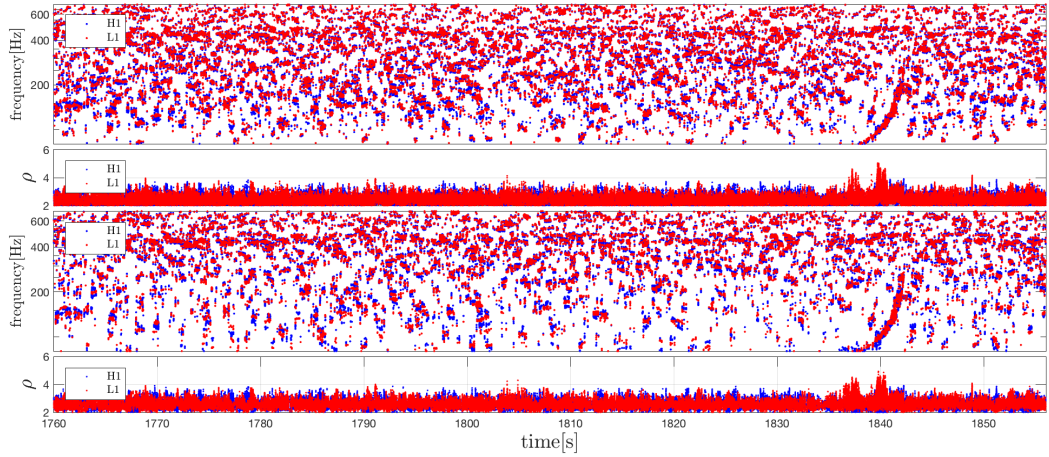

**Figure 1.** (H1,L1)-spectrograms covering GW170817/GRB170817A by coincidences in frequency ($\Delta f = |f_1 - f_2| < 10$ Hz, upper panels) and amplitude ($\rho_1 \rho_2 > 6$, lower panels) of single detector spectrograms of H1 and L1 produced by butterfly filtering. A post-merger feature is apparent in the (H1,L1)-spectrogram merged by frequency more so than when merged by amplitude. Excess amplitudes during GW170817 are primarily due gravitational radiation below 100Hz.

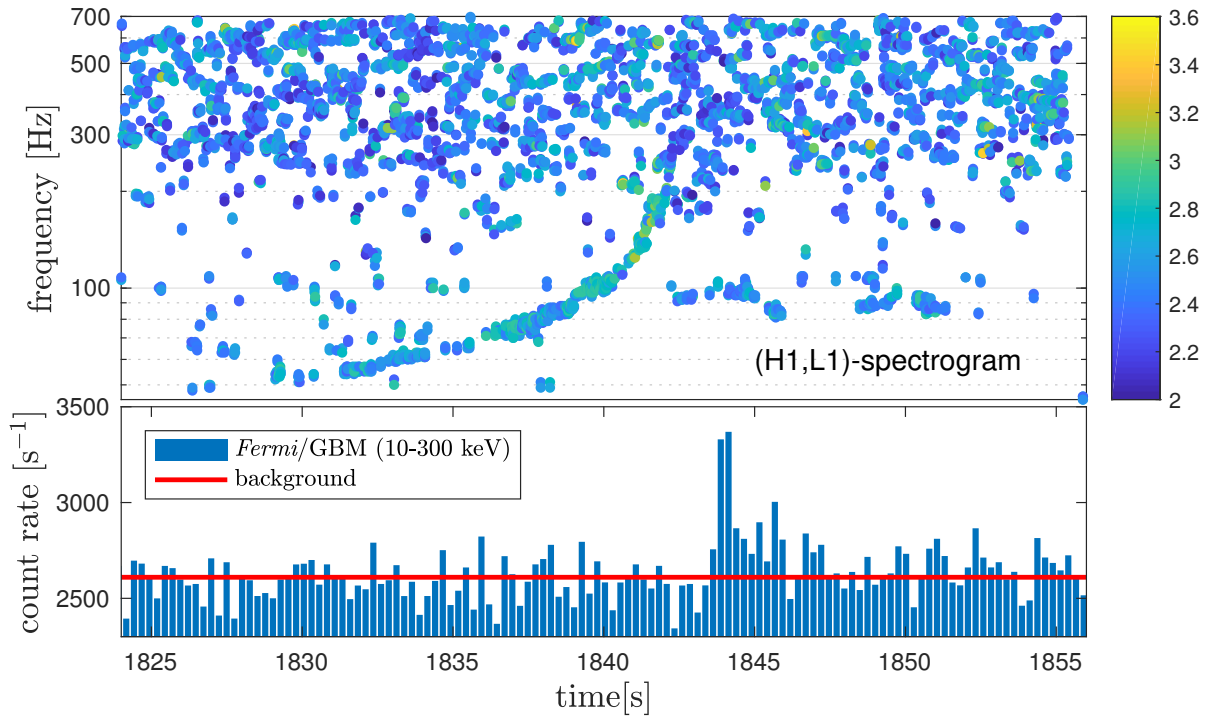

**Figure 2.** Ascending-descending chirp in the (H1,L1)-spectrogram produced by the double neutron star merger GW170817 concurrent with GRB170817A (Goldstein et al. 2017) past coalescence ($t_c = 1842.43$ s). Minor accompanying features around 100Hz (1840-1852 s) are conceivably due to dynamical mass ejecta. Color coding (blue-to-yellow) is proportional to amplitude defined by butterfly output $\rho$ of time-symmetric chirp-like template correlations to data.

For GRB170817A, it appears opportune to search for emissions at energies $E_{gw} \sim 1\% M_\odot c^2$ up to about 1 kHz (Abbott et al. 2017d) given its fortuitous proximity of about 41 Mpc (Cantiello et al. 2018). By causality, Extended Emission in gravitational radiation is potentially meaningful to the central engine if it starts before GRB170817A, that is, during the 1.7 s gap between the time-of-coalescence $t_c = 1842.43$ s inferred from LIGO observations and the onset of GRB170817A detected by *Fermi*-GBM (Connaughton 2017) and INTEGRAL (Savchencko et al. 2017).

Here, we report on a search for broadband extended gravitational-wave emission (BEGE) post-merger to GW170817A in a revisit of LIGO H1, L1 and V1 data from the LOSC (Vallisneri et al. 2014). Our model-independent search uses matched filtering over chirp-like templates that are time-symmetric, allowing detection of ascending and descending chirps with phase-coherence on time scales of order of $\tau$. We use $\tau = 0.5$ s, characteristic for extreme transients gradually exhausting a central energy reservoir over several seconds or more. Fig. 1 shows spectrograms up to 700 Hz obtained by matched filtering over a dense bank of templates demonstrated previously in the identification of broadband Kolmogorov spectra up to 1 kHz in light curves of long GRBs of the *BeppoSAX* catalogue (van Putten et al. 2014).

When the template bank is sufficiently dense, sensitivity approaches that of ideal matched filtering (van Putten et al. 2014; van Putten 2017c), by linear amplification of signals to signal-to-noise ratios greater than unity before seeking coincidences or correlations between two or more detectors. For small signals, sensitivity of this approach appears to exceed that of power excess methods considered in existing model-independent searches for post-merger signals (Abbott et al. 2017).

Application of our *butterfly filter* to LIGO-Virgo data is made possible on a heterogeneous computing platform with multiple high-end *graphics processor units* (GPUs). Single-detector spectrograms are merged in searches for coincident features, that may be identified by edge detection (Canny 1986; MatLab 2018). Gaussian equivalent levels of confidence derive from causality following edge detection background analysis (Supplementary Information).

## 2 OBSERVATION AND INTERPRETATION

Zoomed in to the epoch about GW170817 (Fig. 2), the (H1,L1)-spectrogram shows the relatively loud merger signal of GW170817 with binary coalescence at $t_c = 1842.43$ (Abbott et al. 2017a) described by an *ascending chirp* in gravitational-wave frequency

$$f_m(t) = A(t_c - t)^{-\frac{3}{8}} \quad (t < t_c) \tag{1}$$

identified up to 260 Hz, where $A \simeq 138\,\mathrm{s}^{-5/8}$ Hz is representative of the chirp mass $\mathcal{M} = c(15/768 F(e))^{\frac{3}{5}} (\pi A)^{-\frac{8}{5}} \simeq 1.1382 F(e)^{-\frac{3}{5}} M_\odot$, including $F(e)$ accounting for ellipticity $e \simeq 0$ (Peters & Mathews 1963; Abbott et al. 2017a). The gravitational-wave luminosity during the merger satisfies $L_{GW} = (32/5) (\mathcal{M}\omega)^{\frac{10}{3}} L_0$, where $\omega = \pi f$ and $L_0 = c^5/G$, where $G$ is Newton's constant. In the observed track up to 260 Hz, $L_{GW} \simeq 1.35 \times 10^{50}\,\mathrm{erg\,s^{-1}} \simeq 7.5 \times 10^{-5} M_\odot c^2\,\mathrm{s}^{-1}$, i.e., $4 \times 10^{-10} L_0$, which is relatively gentle compared to $10^{-5} L_0$ of GW150914 at similar frequency.

Fig. 2 shows a continuation of GW170817 with a long-duration *descending chirp during* GRB170817A. It can be fitted by an exponential track (Supplementary Information)

$$f_p(t) = (f_s - f_0)e^{-(t-t_s)/\tau_s} + f_0 \quad (t > t_s) \tag{2}$$

with $\tau_s = 3.01 \pm 0.2$ s, $t_s = 1843.1$ s, $f_s = 650$ Hz and $f_0 = 98$ Hz. Crucially, $t_s - t_c = 0.67$ s shows the descending chirp to start in the 1.7 s gap between GW170817 and GRB170817A, satisfying the aforementioned causality condition. A Gaussian equivalent level of confidence greater than $3.3\sigma$ obtains for significance to the central engine of GRB170817A based on causality over 1952 s of LIGO data. Improved estimates obtain upon including the statistical significance of the strength of the signal (Supplementary Information).

The estimated initial frequency $f_c = 774$ Hz at the time-of-coalescence inferred from (2) is below the orbital frequency at which the stars approach the *Inner Most Stable Circular Orbit* (ISCO) of the system as a whole, i.e., $\sim 1100$ Hz at $\sim 16$ km according to the Kerr metric (Kerr 1963; Bardeen et al. 1972; Shapiro & Teukolsky 1983). At this point, a binary system of two equal mass neutron stars would have a rotational energy $E_{rot} = 4.56\% M_\odot c^2$ and a dimensionless specific angular momentum $\hat{a} = 0.72 < 1$. While this allows prompt collapse to a $\sim 3 M_\odot$ Kerr black hole (Kerr 1963) any gravitational-radiation from remaining debris orbiting about the ISCO would be above 2 kHz. Instead, the $< 1$kHz descending chirp in Fig. 2 points to radiation of $E_{GW} \simeq 0.2\% M_\odot c^2$ from a long-lived rapidly rotating hyper-massive neutron star or magnetar with a dimensionless quadrupole moment of $0.2\%(R/16\mathrm{km})^5$, where $R$ is the radius of the star, induced by dynamical and secular instabilities (Owen et al. 1998; Kokkotas 2008; Corsi & Mészáros 2009) that may include magnetic fields (Abbott et al. 2017d; Cutler 2002)

Magnetar spin down indicated by the descending chirp may include magnetic interactions with surrounding matter from dynamical mass ejections from the preceding merger, in addition to gravitational radiation losses alone. These ejecta are commonly found in aforementioned numerical simulations. The interactions may result from magnetic coupling (Gosh et al. 1977; Gosh & Lamb 1978, 1979) and/or tidal stresses (Hotokezaka et al. 2013) that may include magnetic outflows from open field lines (Lovelace et al. 1995; Parfrey et al. 2016) and gravitational wave emission from quadrupole moments in orbiting debris.

While the descending chirp identifies the central energy reservoir of GRB170817A to be a rotating hyper-massive neutron star or magnetar, its connection to GRB170817A remains uncertain. Perhaps the relatively soft second gamma-ray pulse of GRB170817A (Pozanenko et al. 2018) derives from outflows from debris orbiting the star, while the short-hard pulse derives from an initial outflow from the star at birth. More likely, the complex spectral evolution of GRB170817A derives from shock break-out (Kasliwal 2017; Gottlieb 2018), producing a short-hard pulse followed by softer emission of relatively longer duration as the shock becomes more spherical as it propagates downstream.

## 3 CONCLUSIONS

We present observational evidence for a descending chirp for the first five seconds post-merger to GW170817 (Fig. 2). By frequency, it potentially indicates a magnetar as the central engine of GRB170817A, well below the minimum of 2 kHz emission from high density matter about the ISCO of a $3M_\odot$ black hole. The ultimate fate of the magnetar is uncertain, whether it survives as a pulsar with a spin frequency of 49 Hz or collapses to a black hole at a later stage. The physical mechanism by which the magnetar is protected against prompt collapse is not well understood, but lifetimes of ten seconds such as observed here have been anticipated for proto-magnetars (Ravi & Lasky 2014). Our observation of an extended lifetime appears particularly reasonable in light of the recent LIGO determination of a relatively low total mass of $2.73^{+0.04}_{-0.01} M_\odot$ of the progenitor binary (Abbott et al. 2018) that is just 20% above the neutron star mass of $2.27^{+0.17}_{-0.15} M_\odot$ in PSR J2215+5135 (Linares et al. 2018) and on par with the super-massive PSR J1748-2021B (Freire et al. 2008; Özel & Freire 2016). Future observations promise to significantly improve on these initial observations, and to determine to what extend GRB170817A is representative for canonical SGRBs.

**Acknowledgements.** The authors gratefully acknowledge constructive comments from the reviewer on both the main text and Supplement and stimulating discussions with A. Levinson, B. Barish, A.J. Weinstein, M. Branchesi, E. Chassande-Mottin, G. Prodi, W. Del Pozzo, J. Kanner, R. Weiss, A. Lazzarini, E. Kuulkers and C. Hong, and thank A.M. Goldstein for kindly providing the *Fermi*/GBM light curve in Fig. 2. Support is acknowledged from the National Research Foundation of Korea under grants 2015R1D1A1A01059793, 2016R1A5A1013277 and 2018044640. This work made use of LIGO O2 data from the LIGO Open Science Center provided by the LIGO Laboratory and LIGO Scientific Collaboration. LIGO is funded by the U.S. National Science Foundation. Additional support is acknowledged from MEXT, JSPS Leading-edge Research Infrastructure Program, JSPS Grant-in-Aid for Specially Promoted Research 26000005, MEXT Grant-in-Aid for Scientific Research on Innovative Areas 24103005, JSPS Core-to-Core Program, Advanced Research Networks, and the joint research program of the Institute for Cosmic Ray Research.

**APPENDIX A:**

This Appendix describes a search pipeline for broadband extended gravitational-wave emission comprising (i) whitening, (ii) generating single detector spectrograms by GPU-accelerated butterfly filtering, (iii) merging spectrograms by coincidences in frequency or amplitude and (iv) searches for Extended Emission by image analysis of merged spectrograms. For GW170817 (H1,L1)-data over $T = 1952$ s, the latter comprises a four-dimensional scan for descending chirps by $2.5 \times 10^8$ samples, parameterized by a characteristic time scale $\tau_s$, an initial frequency $f_s$ and a late-time asymptotic frequency $f_0$. Extended Emission (EE) of about 5 s with $\tau_s \simeq (3.01 \pm 0.2)$s is found to start in the 1.7 s gap between GW170817 and GRB170817A. A strict lower bound for the Gaussian equivalent level of confidence derives from *timing on-source:* $3.3\sigma$ by its time-of-onset in the 1.7 s gap. Additional confidence derives from its signal-to-noise ratio SNR of 5.74 above background defined by data outside the 1.7 s gap.

## A1 Introduction

Extreme transient events such as GRBs and core-collapse supernovae may produce broadband extended gravitational-wave emission (BEGE). Time scales of seconds to tens of seconds may be representative for central engines exhausting an energy reservoir in angular momentum. Deep searches for their gravitational-wave emission can be pursued by matched filtering at intermediate time scales of phase-coherence $\tau$, greater than the period $1/f$ of the gravitational wave at frequency $f$ while smaller than the total duration $T$ of the emission. In un-modeled searches, $\tau$ is determined empirically as part of the search.

For GW170817A/GRB170817A, we perform a model-independent deep search for broadband extended gravitational-wave emission in 2048 s (LIGO 2017) of data at 4096 Hz according to Fig. A1 comprising

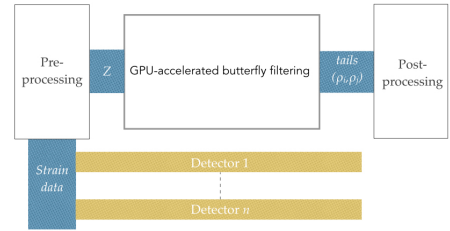

**Figure A1.** Schematic overview of the search pipeline comprising pre-processing and butterfly filtering applied independently to strain data of multiple gravitational-wave detectors. Post-processing produces single detector spectrograms from butterfly output (*tails* of $\rho_i$ and $\rho_j$). Merged spectrograms by coincidences in frequency or amplitude can analyzed for long duration features by $\chi$-image analysis.

- *Pre-processing*: cleaning and glitch removal (Abbott et al. 2017a) followed by whitening of H1, L1 and V1 data;
- *Singe detector spectrograms* by GPU-accelerated butterfly filtering of H1, L1 and V1 by matched filtering over a dense bank of time-symmetric chirp-like templates (van Putten et al. 2014; van Putten 2017);
- *Merging spectrograms* by coincidences in frequency or amplitude, producing merged spectrograms as input to image analysis (van Putten 2018).

Potential Extended Emission is marked by a global maximum $\chi^*$ of an indicator function $\chi$, here of a descending chirp in the form of an exponential feature in merged (H1,L1)-spectrograms (§4). Following LIGO language (e.g. Abbott et al. 2017b), potential significance to the central engine of GRB170817A requires $\chi^*$ to appear with a start time $t_s$ in the gap - the *on-source* window - of 1.7 s between GW170817 and GRB170817A. Over the *background time* $T = 1952$ s set by the length of our data segment, this gives

$$p = \text{FAR} \times 1.7\,\text{s} < 8.7 \times 10^{-3} \qquad (A1)$$

given the *False Alarm Rate* FAR $< 1/T$. Satisfying causality alone hereby implies a Gaussian equivalent level of confidence better than $3.33\sigma$.

Here, we give a detailed account of our analysis over the segment of duration $T$, that highlights the observed time and signal strength $(t_s, \chi^*)$ marking the onset of the Extended Emission in the 1.7 s gap. In this process, background is defined by all data outside this gap.

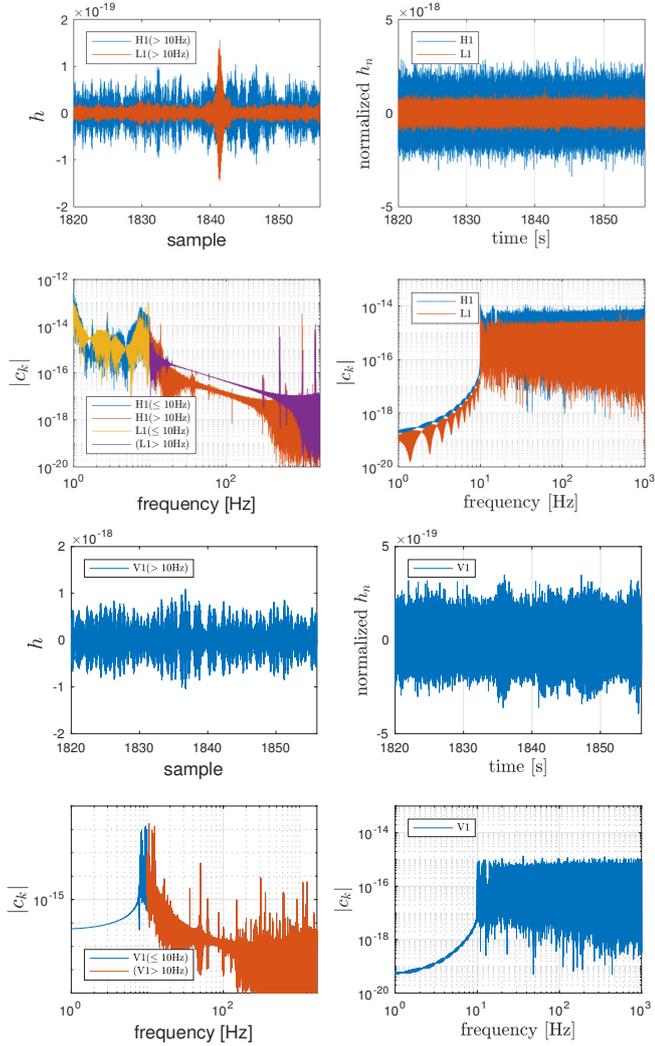

**Figure A2.** (Left panels.) Whitening of H1 and L1 strain data (cleaned data, left panels) by normalisation of standard deviation over segments of 4096 Fourier coefficients $c_k$ (arbitrary units) and a bandpass 10-2000 Hz (right panels). Before normalization, L1 shows a residual following partial removal of a major glitch (LIGO 2017), while violin modes are apparent in the Fourier domain. Both features are suppressed by normalization while preserving GW170817. Note that above $10^1$ Hz, H1 detector noise is slightly below that of L1. (Right panels.) The same whitening is applied to Virgo.

## A2 Whitening

Whitening of strain data is performed by normalization in Fourier domain by amplitude, preserving phase, over a bandpass 10-2000 Hz, here over segments of 4096 coefficients $c_k$ covering bands of 2 Hz. This procedure effectively suppresses narrow band violin modes in the LIGO detectors (Fig. A1), producing essentially uniform spectra while preserving signals of interest. In fact, a further conversion to audio renders GW170817 directly audible.

In the present application to LIGO O2 data covering GW170817, observations of potential interest have been verified to remain unchanged with whitening over different frequency bandwidths of 2-8 Hz.

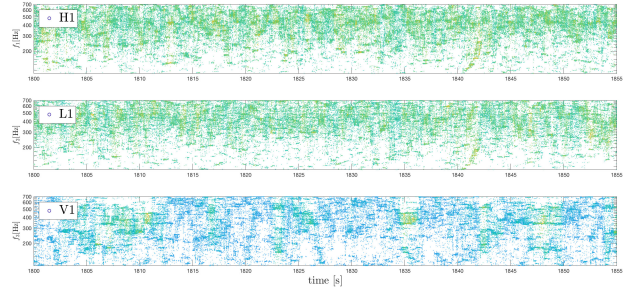

**Figure A3.** Single channel H1-, L1- and V1-spectrograms of butterfly tails $\rho > \kappa\sigma$ ($\kappa = 2$). The majority of data-points are scattered broadly in frequency $f_i$ and $\rho_i$ ($i = 1, 2, 3$). In H1 and L1 (but not V1) GW170817 is visible by clustering of hits, color coded according to $\rho$.

## A3 Single detector spectrograms

Spectrograms are generated by matched filtering against a dense bank of templates of intermediate duration $\tau$, that covers a broad range in frequency $f(t) < f_1$ and finite time rate-of-change of frequency

$$0 < \delta \leqslant \left|\frac{\tau df(t)}{f dt}\right|, \qquad (A2)$$

that defines our butterfly filtering (van Putten et al. 2014; van Putten 2017). In the present search, single detector spectrograms are generated over frequencies up to $f_1 = 700$ Hz using time-symmetric templates of duration $\tau = 0.5$ s, whose frequencies vary slowly in time ($0.1\% < f^{-1}\Delta f < 10\%$). At 500 Hz, for example, matched filtering with phase-coherence over $\tau = 0.5$ lifts a signal by a factor $\sqrt{N} \simeq 11$, where $N$ is the number of periods over $\tau$ (van Putten 2016).

A deep search for un-modeled signals densely covering a region of $f(t)$ and $df(t)/dt$ is realized by bank of up to millions of templates, efficiently processed on heterogeneous computing platform with *graphics processor units* (GPUs) (van Putten 2017).

In butterfly filtering, potential signals are lifted to signal-to-noise ratios greater than one in tails $\rho_i$ of H1, L1 and V1 ($i = 1, 2, 3$), here with $\kappa = 2$ (Fig. A3). Tails from H1 and L1 are essentially uncorrelated in fluctuations in frequency and amplitude (Fig, A4), measured by Pearson coefficients

$$\mathcal{C}_{12} = \begin{pmatrix} C_{f_1 f_2} & C_{f_1 \rho_2} \\ C_{f_2 \rho_1} & C_{\rho_1 \rho_2} \end{pmatrix} = \begin{pmatrix} -0.0007 & -0.0015 \\ -0.0029 & 0.0002 \end{pmatrix}. \quad (A3)$$

This permits searches for potential signals by pairing, i.e., features in (H1,L1)-spectrograms merged by coincidences in frequency or amplitude $\rho(f)$. It is hereby meaningful to insist on candidate signals to be detected in spectrograms merged by frequency or amplitude.

Butterfly filtering has previously been applied to time-series of gamma-rays from long GRBs of *BeppoSAX* (van Putten et al. 2014) and LIGO S6 van Putten (2017). In the first, it identifies a broadband Kolmogorov spectrum up to 1 kHz in light curves with a mean of 1.26 photons/0.5 ms in the first. The second enabled detailed sensitivity tests against software and LIGO hardware injections, showing a regular increase in total number of hits in tails of butter-

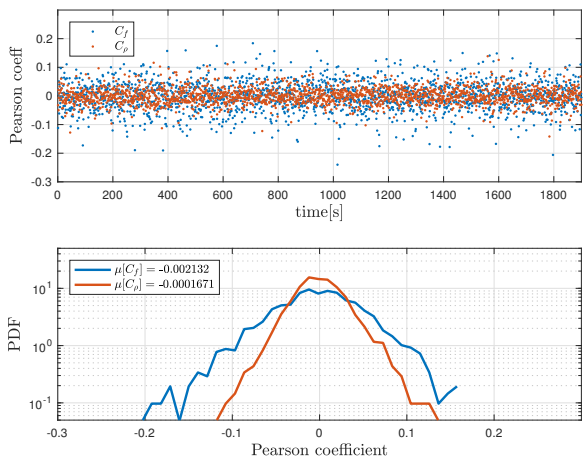

**Figure A4.** Correlations $C_f$ and $C_\rho$ in fluctuations in frequency and, respectively, $\rho$ in across H1 and L1 single detector butterfly output (upper panel). The results show an essentially correlation-free background (lower panel). Results shown are from a test over a bank of 32 k chirp templates broadly covering 100-700 Hz.

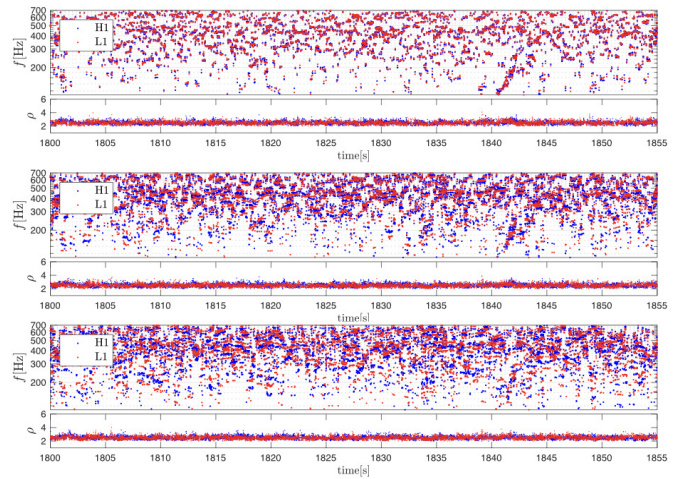

**Figure A5.** Merged spectrograms by small frequency differences $\Delta f = |f_1 - f_2| \leqslant 12\,\text{Hz}$ ($f_1$ from H1, $f_2$ from L1) brings about GW170817 along with a continuation in the form of a descending chirp (top panel). The ascending-descending chirp of GW170817/GRB170817A seen is captured by chirp-templates with bandwidths up to tens of Hz. It satisfies causality, in fading out upon merging with frequency differences greater than $14\,\text{Hz} \leqslant \Delta f \leqslant 28\,\text{Hz}$ (middle panel) and $28\,\text{Hz} \leqslant \Delta f \leqslant 42\,\text{Hz}$ (lower panel).

fly output with signal injection strength. Parameterized by search depth $\kappa$, these tails are defined by

$$\rho_{ij}(t) > \kappa \sigma_{ij} \quad (A4)$$

where $i$ denotes the template index and $\sigma_{ij}$ is the standard deviation in correlations $\rho_{ij}(t)$ between template $i$ and a data segment $j$ over time $t$. The bulk of the output $\rho_{ij}(t)$ in $t$ is essentially Gaussian, and any candidate signal appears in non-Gaussian behavior of (A4). In what follows, we drop subscripts $ij$ in reference to (A4).

In the present application to GW170817, observations of potential interest have been verified to remain unchanged with small variations in $\tau$. Major variations in $\tau$ can result in capturing different signals, as increasing $\tau$ increases (decreases) sensitivity to signals with phase-coherent over time scales $> \tau$ ($< \tau$).

### A4 Merging spectrograms

Single detector spectrograms can be merged by small frequency difference coincidences $\Delta f = |f_1 - f_2|$ between pairs of hits or a lower bound on the amplitude $\rho_1 \rho_2$ (Fig. 1, main text). Merging reduces the total number of data-points and potential signals will appear in clustering of hits.

In what follows, we focus on (H1,L1)-spectrograms merged by frequency, since this appears to be relatively more sensitive than (H1,L1)-spectrograms merged by amplitude. This is perhaps not surprising by analogy to the typically superior performance of *frequency modulation* (FM) over *amplitude modulation* (AM) in in the face of interferences of manmade and natural origin.

Discrete confirmations of candidate signals derive from (i) causality, i.e., the onset of a post-merger feature potentially relevant to the central engine must be in the 1.7 s gap that, furthermore, should fade out in merging by frequency differences $\Delta f$ *greater* than some positive lower bound; and (ii) an accompanying imprint in spectrograms merged by amplitude. Quantitative assessment derives from signal-to-noise ratio (SNR) and background image analysis applied to such merged (H1,L1)-spectrograms.

#### A4.1 (H1,L1)-spectrograms merged by frequency

For chirp-like signals with finite $df/ft > \delta$ for some $\delta > 0$, differences in frequency in H1 and L1 arise from $\delta t \times df/ft$, where $\delta t < 10\,\text{ms}$ is the difference in time-of-arrival at the two detectors, and approximate matches in matched filtering against chirp-like templates with finite bandwidths.

Searching for broadband extended gravitational-wave emission, individual spectrograms are merged by passing through hits with small differences $\Delta f$, here with $\Delta f$ less than about 10 Hz (Fig. A5).

Candidate features appear in (H1,L1)-spectrograms by a clustering of hits (A4), more so than exceptional values of $\rho = \rho_{ij}$ of a particular template $i$ and data segment $j$. Candidate features readily obtain by Canny edge detection (Canny 1986). For the ascending-descending feature covering GW170817/GRB170817A, we verify that this extended feature fades out in merged spectrograms with increasing frequency differences $\Delta f$ in equal time H1 and L1 single detector spectrograms (Fig A5).

As Extended Emission, the descending chirp feature is of particular interest since it starts in the 1.7 s gap between GW170817 and GRB170817A. We next turn to an indicator function to quantify its potential significance.

#### A4.2 An indicator function $\chi$

For the post-merger descending chirp in the (H1,L1)-spectrogram (Fig. A5), we mark such features by count-ing the number $c$ of hits in time-frequency bins (0.1 s × 10 Hz) along "time-frequency tracks" (e.g Thrane & Coughlin 2014), here the exponential curves (Figs. A6-A7)

$$f_p(t) = (f_s - f_0)e^{-(t-t_s)/\tau_s} + f_0 \quad (A5)$$

over extended time intervals $t_s < t < t_s + \Delta t$ ($\Delta t = 7$ s) following a start time $t_s$, parameterized by a characteristic time scale $\tau_s$ and an initial frequency $f_s$ and a late-time frequency $f_0$. Following normalization of $c$, we define the *indicator*

$$\chi = \frac{c - \mu}{\sigma}, \quad (A6)$$

where $\mu$ and $\sigma$ refer to mean and standard deviation of $c$ over the two-dimensional region parameter space $(t_s, \tau_s)$. Normalization (A6) renders the probability density function (PDF) of $\chi$ to be approximately Gaussian (below).

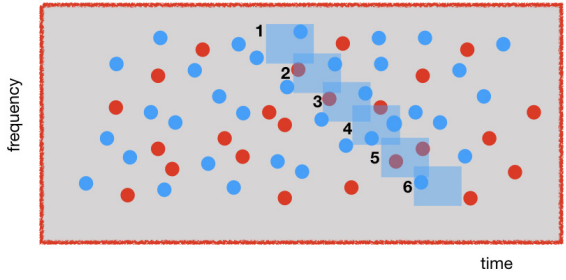

**Figure A6.** Schematic illustration of edge detection by $c = \sum_{i=1}^{6} c_i = 0 + 1 + 1 + 1 + 0 + 0 = 3$, counting the number of coincident hits in a merged spectrogram along a strip, here descending, discretized by cells of width $\delta f$ and $\delta t$ in frequency and time.

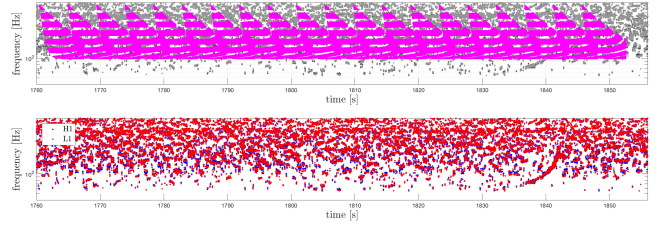

**Figure A7.** Scanning the (H1,L1)-spectrogram by strips partitioned into blocks of size $\delta t \times \delta f$, $\delta t = 0.1$ s, $\delta f = 10$ Hz, along curves from the family (A5) produces counts of hits by H1 *and* L1. These counts $c$ are normalized to $\chi$ (A6). The spectrogram shown is the first frame of 96 s *after* (top panel) and *before* (bottom panel) scanning, as part of a complete scan over $T = 1952$ s.

Fig. A8 shows a global maximum

$$\chi^* = 6.5993 \quad (A7)$$

at start time $t_s = 1843$ s in the 1.7 s gap between GW170817 and GRB170817A with

$$\tau_s \simeq 3.01 \text{ s}, \quad f_s = 650 \text{ Hz}, \quad f_0 = 98 \text{ Hz}. \quad (A8)$$

This maximum is identified in a 3+1 dimensional scan $\chi = \chi(\tau_s, f_s, f_0, t_s)$ comprising

$$N_s = N_1^3 \times N_2 \simeq 250 \times 10^6 \quad (A9)$$

samples through $t_s \epsilon [0, T]$ s ($N_2 \simeq 60,000$ steps $\delta t_s = 0.031$ s). Using $N_1 = 16$, this scan interpolates $\tau_s \epsilon [0.5, 4]$ s with $\delta \tau_s = 0.23$ s and considers the 30 Hz intervals $f_s \epsilon [634, 664]$ Hz and $f_0 \epsilon [84, 114]$ Hz with $\delta f_s = \delta f_0 = 2$ Hz.

Fig. A9 shows the results of the two-dimensional result for $\chi$ with aforementioned $f_s = 650$ Hz and $f_0 = 98$ Hz.

### A4.3 Background statistics of $\chi$

A PDF of background statistics of $\chi$ is obtained from a normalized histogram over bins $\Delta \chi = 0.0846$ with an associated cumulative distribution of $\chi$ defined by sampling data outside the 1.7 s gap (Fig. A8).

The tail of the PDF of $\chi$ is given e Gaussian fit with $\sigma = 1.15$, slightly different from the unit standard deviation in $\chi$ in (A6). The global maximum $\chi^* = 6.553$ hereby satisfies

$$\text{SNR} = 5.7385. \quad (A10)$$

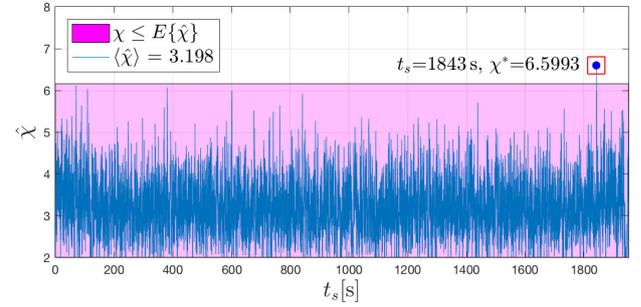

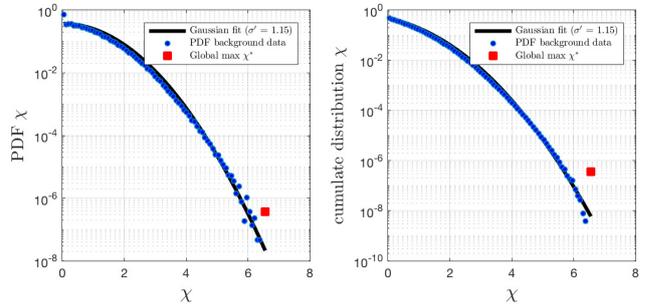

**Figure A8.** (Upper panel.) Shown is $\hat{\chi}(t_s) = \max_{\tau_s, f_s, f_0} \chi(\tau_s, f_s, f_0, t_s)$ and its global maximum $\chi^*$ at $t_s = 1843$ s in the 1.7 s gap between GW170817 and GRB170817A from a 3+1 scan over $(\tau_s, f_s, f_0, t_s)$ comprising $N_s \simeq 250 \times 10^6$ samples. (Lower panels.) Shown are the PDF of background data (*blue dots*) defined by $\chi$ over data excluding the the "on-source" gap of 1.7 s with Gaussian fit to the tail ($\sigma = 1.15$, *black curves*), the associated cumulative distribution and the global maximum $\chi^*$ (*red square*).

The lower panels of Fig. A8 show a probability $p\Delta\chi = 3.8860 \times 10^{-7}$ of $\chi$ in excess of $p(> \chi^*) = \int_{\chi^*}^{\infty} p(x)dx = 5.9384 \times 10^{-8}$ of the cumulative PDF of our Gaussian fit, giving a $p$-value

$$p = 0.016 \quad (A11)$$

with Gaussian equivalent level of significance $2.43\sigma$.

To guide the eye, included in Fig. A8 is the expected range $\chi \leqslant E\{\hat{\chi}\}$ defined by absolute values of samples of size $n$ taken from a Gaussian with standard deviation $\sigma$ (adapted

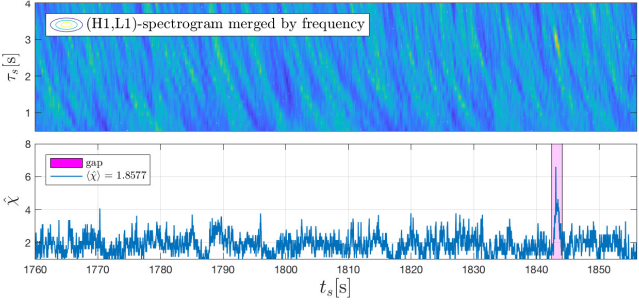

**Figure A9.** (Top two panels.) $\chi(\cdot,\cdot,f_s,f_0)$ for $f_s = 650$ Hz and $f_0 = 98$ Hz that contains that global maximum $\hat{\chi} = 6.5993$ over $T = 1952$ s in scan for exponential tracks (A5) with start time $t_s = 1843.1$ s in the 1.7 s gap between GW170817 and GRB170817A.

from van Putten 2016)

$$p(x) = \frac{n}{\sqrt{2\pi}\sigma}\mathrm{erf}(y)^{n/2-1}e^{-y^2}, \quad y = \frac{x}{\sqrt{2}\sigma} \quad (A12)$$

with expectation values $\bar{x}$ satisfying $n\mathrm{erfc}(\bar{y}) = 1/2$. In Fig. A8, $\hat{\chi} = \hat{\chi}(t_s)$ is defined by maximum values of samples over $(\tau_s, f_s, f_0)$. Based on (A12), the observed time average $\langle\hat{\chi}\rangle \simeq 3.198$ over $t_s$ shows a true number of independent samples $n = 222$ well below $N_1^3 = 4096$ in $(\tau_s, f_s, f_0)$ - our four-dimensional scan is effectively over-sampling the two-dimensional (H1,L1)-spectrogram by a factor of about 18. (A similar result applies to Fig. A9.) On the basis of $n = 222 \times N_2$ and $\sigma = 1.15$, (A12) gives an expectation value $E\{\hat{\chi}\} = 6.1561$.

In the upper panel of Fig. A8, global maximum $\chi^*$ is the peak of three similar values $\hat{\chi} = [6.5993, 6.5525, 6.5575]$ clustering in $1843.1 < t_s < 1843.2$. Relative to $E\{\hat{\chi}\} = 6.1561$, they carry $p$-values $[0.09, 0.13, 0.13]$. Ignoring clustering and adjusting for aforementioned over-sampling, these peak values carry a $p$-value of about

$$p = 0.029 \quad (A13)$$

with corresponding Gaussian equivalent level of significance of $2.2\sigma$. This result is very similar to A11.

Finally, the same $\chi$-image analysis has been applied to spectrograms merged by $\rho$ (Fig. 1, main text), confirming $\chi^*$ at $t_s$ in A9.

### A5  Significance of the global maximum

The outcome of our analysis is a global maximum in $\chi^*$ over the window of $T = 1952$ s with attributes $(t_s, \chi^*)$ (Fig. A8). With an approximately stationary background noise in H1 and L1 over $T = 1952$ s considered here, $(t_s, \chi^*)$ are statistically independent observables. Their respective PDF's are uniform and, respectively, approximately Gaussian (Fig. A8). Consequently, we consider two independent $p$-values as follows:

*(a) $t_s$ satisfies causality* in the outcome Fig. A8) of our image analysis of the (H1,L1)-spectrogram over $T = 1952$ s of data. Specifically, $\chi^*$ satisfies

$$t_s - t_c = 0.67 \text{ s}, \quad (A14)$$

i.e., it falls in the 1.7 s gap between time-of-coalescence $t_c$

in GW170817 and the onset of GRB170817A. This carries $p = 1.7/1952$, i.e.,

$$p = 8.7 \times 10^{-4}: \quad 3.3\sigma \quad (A15)$$

of (A1). Additional confidence derives from confirmation of the same in the (H1,L1)-spectrogram merged by amplitude. This $p$-value can be improved by analysis over a larger dataset from LIGO O1 or O2.

*(b) $\chi^*$ is above background* with SNR = 5.7385 and equivalent $p$-value

$$p \simeq 0.03: \quad 2.2\sigma, \quad (A16)$$

taking the maximum of the estimates (A11) and (A13). Equivalently, this improves on FAR in (A1) to about once per day. This extrapolation may be tested empirically by analysis of more extended data from LIGO O1 or O2. A realistic prospect for improvement is further by enhanced detector sensitivity anticipated with the upcoming LIGO O3 observations, scheduled for 2019.

By the above, the level of confidence of total significance of the descending chirp to the central engine of GRB170817A satisfies

$$p = 2.5 \times 10^{-5}: \quad 4.2\sigma. \quad (A17)$$

**Data and software availability.** The LIGO O2 data covering GW170817/GRB170817A are available from the LOSC (van Putten 2018). Butterfly filtered H1L1-data and the $\chi$-Image Analysis MatLab script are deposited on Zenodo.org (van Putten 2018).

This paper has been typeset from a TeX/ LaTeX file prepared by the author.